\documentstyle[12pt,epsf]{article}
\newcommand{\half}{\mbox{$\frac{1}{2}$}}
\renewcommand{\theequation}
{\arabic{section}.\arabic{equation}}

\begin{document}
\hfill UPR-0085NT\\
\begin{center}
\vspace{2cm}
{\Large \bf Quantum theory of large amplitude collective motion
and the Born-Oppenheimer method}\\
\vspace {1cm}
{\bf Abraham Klein and Niels R.\ Walet}        \\
{\small \it Department of Physics, University of Pennsylvania, \\
Philadelphia, Pennsylvania 19104-6396, USA} \\
\end{center}
\hspace*{20pt}
\begin{center}
{\bf Abstract} \\          \end{center}
{\small
We study the quantum foundations of a theory
of large amplitude collective motion for a Hamiltonian expressed
in terms of canonical variables.  In previous work the separation
into slow and
fast (collective and non-collective) variables was carried out
without the explicit intervention of the Born Oppenheimer
approach.  The addition of the Born Oppenheimer assumption
not only provides
support for the results found previously in leading approximation,
but also facilitates an extension of the theory to include
an approximate description of the fast variables and their interaction
with the slow ones.  Among other corrections, one encounters
the Berry vector and scalar potential.
The formalism is illustrated with the aid of
some simple examples, where the potentials in question are actually
evaluated
and where the accuracy of the Born Oppenheimer approximation is tested.
Variational formulations of both Hamiltonian and Lagrangian type are
described for the equations of motion for the slow variables.}

\section{Introduction}

For the past decade, the authors and their collaborators have
been involved in the development of a theory of large amplitude
collective motion.  This enterprise began with a pair of
papers \cite{K1,K2} on possible quantum foundations for such a program.
Soon thereafter \cite{DK1} it was realized that the leading
approximation,
to which most previous studies had been confined,
was classical in nature, (with subsequent requantization) and,
as a consequence, almost all later
systematic theoretical development, reviewed in Ref.~\cite{KWDR},
was based on
the study of this limit.

Quantum corrections can be important
in selected circumstances,
however.  Consequently, the elements of a systematic method for
including quantum corrections by expansion about the classical limit
has recently been developed \cite{WKDQ}, distinct from the approaches
found in our initial papers.  We consider it worthwhile, nevertheless,
in the light of subsequent developments,
especially within the context of the Berry phase idea \cite{B1,MSW},
to update, augment and improve the work described in those initial
efforts and to tie them, where possible, to more recent work.
In a paper currently in preparation, we shall contrast the
nuclear-physics
foundations of the method described in this paper with that found in
Ref.~\cite{WKDQ}, both approaches requiring elaboration compared
to what
has already been published.

In Sec.\ \ref{sec:oldstuff} we give a condensed but
at the same time more precise account of the essential theoretical
content of Ref.~\cite{K1}, namely, a method for separating and
identifying
a collective subspace for a specified class of Hamiltonians.
This is done in an approximation that suppresses any coupling
to non-collective degrees of freedom.  As opposed to the corresponding
problem in molecular physics, the Hamiltonian is generally
given in a form where the coordinates are not appropriate to make the
separation of the Hamiltonian into a collective and a non-collective
part.
The discovery of new
coordinates that will effect the separation is an essential part of the
problem.  In the regime of large amplitude collective motion, a
reasonably complete theory that can be implemented \cite{KWDR,WKDQ}
has been developed
so far only for the case that the transformations to new coordinates
and momenta are restricted to point transformations.

In Sec.\ \ref{sec:BO}, we describe how the previous theory may be
extended to
include the interaction between the fast and slow variables.
This is done by means of
a standard Born-Oppenheimer (BO) representation of the states of the
slow variables.   Though some of the ensuing details are similar to
those encountered in the molecular problem, leading, for example,
to the
occurrence of Berry potentials \cite{B1,MSW}, others are characteristic
of the problem of large amplitude collective motion.  In order to
evaluate the corrections found to the potential energy of the
collective variables, it is necessary to obtain wave functions
for the fast variables.  This is done by an
extension of the analysis carried out in Sec.\ \ref{sec:oldstuff}.
At the same time
this development provides a justification for
some of the assumptions of the previous analysis.

In Sec.\ \ref{sec:varia}, we present material that has no direct
counterpart
in any previous work
in this field, outside our own.  One of the results of
Sec.\ \ref{sec:BO} was to establish a complete effective
quantum mechanics in the
collective subspace, i.\ e., we not only computed an effective
Hamiltonian operator in the collective subspace, but also proved
that it was expressed in terms of canonical variables.  In
Sec.\ \ref{sec:varia}
we study variational principles for the associated Heisenberg matrix
mechanics. If we put aside, momentarily, special difficulties
associated
with the curved spaces that we allow in our general formulation, we
find quantum analogues of variational principles both for Hamilton's
and for Lagrange's equations.  The connection between the two forms
is the standard one, even when Berry phase terms are included, in
contrast to a result based on the use of path integrals \cite{MSW}.
For the general case of a curved space the situation is more involved,
and only a variational principle for Hamilton's equations is presented
and analyzed.

The corrections to the potential energy of collective motion
found in Sec.\ \ref{sec:BO} are of three types.  There is first a
``quantum"
potential arising from the curved space (position-dependent mass
tensor).  This term cannot be evaluated fully with the tools developed
in our previous work \cite{KWDR}.  In practical applications to
systems with
$N$ degrees of freedom, this term is of order $N^{-2}$,
and therefore should be small for a
many-body system.  A second term consists of the eigenvalue for the
state of the fast variables for a fixed value of the slow variables.
In our approximation this state is that of a set of independent
harmonic oscillators, with frequencies dependent on the instantaneous
value of
the collective  coordinates.
We are mainly (though not exclusively) interested in the zero-point
motion associated with the ground state.  This contribution, which is
only of order $N^{-1}$ compared to the leading term has been studied
previously in simple examples \cite{WKDC}.
A third correction, referred to as the Berry scalar potential,
occurs together with the Berry vector
potential, both appearing in the collective Hamiltonian as a
consequence of
the BO approximation.

Examples in the literature relevant to the approach described in this
paper are not numerous.  The examples studied by Bulgac \cite{Bu1,Bu2}
have been stimulating, but a bit too simple for our purposes.
On the other     hand, the work by
Girard, LeTourneux, and Vinet \cite{GTV} on the dipole-quadrupole
problem
in spherical nuclei is too complex to satisfy our initial needs.
Sec.\ \ref{sec:models} is devoted to the study of several
``in-between" models sufficiently
simple that the BO approximation can be carried out,
including the calculation of the Berry potentials, and compared with
more
exact results.
Some technical details associated with the material of
Secs.\ \ref{sec:oldstuff} and \ref{sec:BO},
respectively, are provided in two appendices.

\setcounter{equation}{0}
\section{Summary of previous results \label{sec:oldstuff}}

In this paper we base our study on a Hamiltonian, that with the help
of the
summation convention, takes the form
\begin{equation}
H = \tilde{H}(\xi^\alpha,\pi_\alpha)=\frac{1}{8} \{\pi_\alpha,
\{\pi_\beta,\tilde{B}^{\alpha\beta}(\xi)\}\} + V(\xi),
\label{eq:2.1}       \end{equation}
that describes $N$ coordinate and momentum pairs,
$\xi^\alpha$ and $\pi_\alpha$, $\alpha=1...N$, that satisfy canonical
commutation relations, ($\hbar=1$),
\begin{equation}
[\xi^{\alpha},\pi_{\beta}] = i\delta^\alpha_\beta.  \label{eq:2.100}
\end{equation}
We thus allow for a curved space described by a (reciprocal)
mass tensor $\tilde{B}^{\alpha\beta}$.  Not only is this formulation
of interest for a range of applications outside of nuclear physics
\cite{KWDR,WKDC}, but it is also possible \cite{KM1}
to transform
the large amplitude problem of nuclear collective motion into this form
as the basis for further development. Associated with this
Hamiltonian is a form for the scalar product in Hilbert space
that is discussed in appendix A.

The problem  of immediate
interest is to find a decomposition of the operator
${\tilde H}$ into two parts, exactly in rare cases, approximately
under most circumstances, that describe slow and fast
(or collective and
non-collective) degrees of freedom.  A basic difficulty of the class
of problems that interest us is that we cannot assume that this
separation occurs for the initial choice of coordinates.
Instead, we assume that it is possible to effect
such a decomposition, at least approximately, by means of
a point transformation of the form
\begin{eqnarray}
   \xi^\alpha &=& g^\alpha(x), \nonumber \\
    x&= & \{Q^i ,q^a\}, \nonumber   \\
   i&=&1...K, \nonumber \\
   a&=&K+1...N,  \label{eq:2.2}             \end{eqnarray}
that is locally invertible,           \begin{equation}
x^\mu = f^\mu(\xi).   \label{eq:2.3}            \end{equation}
Though we have discussed, in the past \cite{DK1,KWDR},
the possible interest
of going beyond point transformations in the study of large amplitude
collective motion, it is only for this simpler class of
transformations that a substantial theoretical underpinning can be
claimed to exist, a foundation that we wish to widen in the present
work.
(Of course, there is a substantial literature based on transformations
that are polynomial in coordinates and momenta that are applicable
to anharmonic vibrations (see, for instance \cite{YK1,SMMZ}), but
this is not the subject of the current paper.)

A first step is to carry out a formal transformation of the
Hamiltonian,
$H$, to the new variables.  For the potential energy, we have,
trivially enough,                      \begin{equation}
{\tilde V}(\xi)={\tilde V}(g^\alpha(Q,q))\equiv V(Q,q). \label{eq:1.1}
                                      \end{equation}
Turning then to the consideration of the kinetic energy, ${\tilde T}$,
\begin{equation}
{\tilde T}=\frac{1}{8}\{\pi_\alpha ,\{\pi_\beta ,
\tilde{B}^{\alpha\beta}
(\xi)\}\},        \label{eq:2.16}             \end{equation}
this expression can be transformed to the new variables with the aid
of the relations \cite{K1}
\begin{eqnarray}
\pi_\alpha&=& \half\{f^\mu_{,\alpha},p_\mu\},\label{eq:2.17} \\
B^{\mu\nu}&=& f^\mu_{,\alpha}\tilde{B}^{\alpha\beta}f^\nu_{,\beta}.
\label{eq:2.18}                      \end{eqnarray}
The derivation of (\ref{eq:2.17}) is given in appendix A, whereas
(\ref{eq:2.18}) confirms the tensorial character of the mass tensor.
In terms of the new variables, the kinetic energy
consequently takes the form               \begin{equation}
T=\frac{1}{8}\{p_\mu ,\{p_\nu ,B^{\mu\nu}(Q,q)\}\} + U(Q,q),
\label{eq:2.19}                       \end{equation}
where the second term, which is specifically a quantum potential
arising from
the non-commutativity of coordinates and momenta, has the form
\begin{equation}
8U(Q,q)=[f^\mu_{,\alpha\gamma}g^\gamma_{,\nu}g^\alpha_{,\lambda}
B^{\nu\lambda}]_{,\mu} - [f^\mu_{,\alpha\gamma}g^\gamma_{,\mu}]_{,\nu}
g^\alpha_{,\lambda}B^{\nu\lambda}.  \label{eq:1.2}  \end{equation}
For further work, the momenta are also divided
into collective and non-collective subsets, \begin{equation}
p_\mu = \{P_i,p_a\}.               \end{equation}

We now remind ourselves that the initial, even
primary, goal of the present
considerations is to identify a piece of the transformed Hamiltonian
as collective.  This collective Hamiltonian, $H_C$, should depend only
on the variables $Q^i,P_i$, that we suppose to be canonical pairs.
The simplest possible choice would appear to be the restriction of
the full Hamiltonian to the values $q^a=0,\;p_a=0$, and thus (see
below)
                     \begin{equation}
H_C\cong \frac{1}{8}\{P_i,\{P_j,B^{ij}(Q,0)\}\} +V(Q,0).
\label{eq:1.3}              \end{equation}
This choice is purely formal until we specify a procedure for
determining the unknown functions $V(Q,0)\equiv V(Q)$ and $B^{ij}(Q,0)
\equiv B^{ij}(Q)$ in terms of the elements of the original Hamiltonian
and of the point transformation (\ref{eq:2.2}) and (\ref{eq:2.3}).  The
criteria for such a determination must be chosen so as to somehow
minimize the coupling between the collective and non-collective
variables.

The simplest approach to this problem is to consider $H_C$,
(\ref{eq:1.3}), as the first term of a Taylor expansion in powers
of the non-collective variables, $q^a$, remembering at the same time
that we have a polynomial of degree two in the non-collective momentum
operators, $p_a$.  Assuming that higher order terms are successively
of lesser importance, basically the same assumption that governs the
treatment of all variables in the small vibrations domain,
we adopt the same assumption that defines optimum decoupling
in the classical limit \cite{KWDR}, namely
that the terms linear in the non-collective
variables vanish.  This yields the two sets of conditions
                             \begin{eqnarray}
V_{,a}(Q) +\frac{1}{8}\{P_i,\{P_j,B_{,a}^{ij}(Q)\}\}&=&0,
\label{eq:1.4}\\
B^{ai}(Q) &=& 0,    \label{eq:1.5}    \end{eqnarray}
recognizable as the coefficients of $q^a$ and $p_a$, respectively.
The operator character of these conditions is apparent only from
the second term of (\ref{eq:1.4}).  However, if we introduce
the Wigner transform of these equations, because of the special
symmetrization that we have adopted for the kinetic energy, they
reduce exactly to the classical conditions  studied
extensively in our previous work
\cite{KWDR}.  There, starting from
these classical conditions, we have described and illustrated
suitable algorithms for the determination of the elements of the
collective Hamiltonian.

With the determination of the elements of the Hamiltonian,
(\ref{eq:1.3}),
the theory is completed by the assumption that the eigenstates of this
operator are defined on a K-dimensional Hilbert space of
(round-bracketed) states $|n)$,   \begin{equation}
|n)=\int dQ |Q) (Q|n),     \label{eq:1.6}    \end{equation}
and that these states accurately model a corresponding subspace,
$|n\rangle$,
of states of the full Hilbert space.  The theory described above
is very closely
the theory that has been applied in a number of early applications,
such as \cite{UK,BKD}. A major aim of the present work is to critique
and
generalize this procedure.

We emphasize that in this
program, the elements of the collective Hamiltonian
are determined from a K-dimensional manifold in configuration
space, parametrized by the coordinates $\{Q\}$, whereas the quantum
mechanics is determined in a Hilbert space defined by the
same coordinates.
Upon further reflection, it may strike the reader
that this is a strange result, or, at least, a very limited one.
The reason is that the true
collective states, $|n\rangle$, though they define
a subspace of the full Hilbert space, are nevertheless states in
this space
and therefore depend also
on the non-collective coordinates, at the very least through the
zero-point
motion of the latter.
Though we have been able to neglect this dependence in the
approximation
considered so far, this will no longer be true if we seek
to include the effect of the coupling to the
fast variables on the properties of the collective states,
or in directly describing the motion of the fast variables.

\setcounter{equation}{0}
\section{Generalization of the formalism to include the effects
of the fast variables
 \label{sec:BO}}

In this section, we set ourselves two tasks.  We shall first find the
forms of the leading corrections to the collective Hamiltonian
given above, arising
from the coupling of the fast to the slow variables. This will be done
with the help of the standard Born-Oppenheimer (BO) approximation.
In order to evaluate these corrections explicitly, however, we shall
then commit ourselves to further approximations for the
dependence of the full Hamiltonian
on the fast variables, that yield a normal
mode description of the latter.

To generalize the restricted ideas of the previous section,
we thus introduce
the BO picture into the theory of large
amplitude collective motion.  As stated above
this amounts to an extension of the
complexity of structure allowed for the states of the collective
subspace so as to take into account the influence of the fast variables.
We first assume, more generally,  that the collective states
$|n\rangle$ have coordinate space representatives
\begin{eqnarray}
\langle Q^i q^a|n\rangle  &\equiv & \langle Q,q|n\rangle \nonumber \\
        &=&\sum_\nu(Q|n\nu)[q|\nu\!:\!Q],
\label{eq:2.4}                      \end{eqnarray}
where the index $\nu$ is not to be confused with its previous use as a
 coordinate index.
Here and for the remainder of this paper, we adopt a notation where
angular
brackets indicate states in the full Hilbert space, square brackets
states
in the space of fast variables (though dependent parametrically on
the slow
variables. as denoted by $:\!Q$ in the state vector), and parentheses
states    in the collective space.
We suppose that for fixed $Q$, the states $[q|\nu\!:\!Q]$ are a
complete set of functions for the fast coordinates,
\begin{equation}
   \sum_\nu [q|\nu\!:\!Q][\nu\!:\!Q|q']
    = \delta(q-q').     \label{eq:2.5}         \end{equation}
Though for the moment we have not specified the equation
of which they are the solutions, we shall be able to do so, at least
approximately, as a consequence of the developments to be carried out
 in this section.

For the remainder of the current discussion, we shall consider the
simplest case in which the fast variables occupy, for any given value
of the slow variables, their state of lowest energy, which is assumed
to
be non-degenerate.
 For the case to be studied here,
$\nu$ takes a single value denoted by zero.
The coefficient function, usually
 denoted in this case by
$(Q|n)$, that appears in (\ref{eq:2.4}), can then
be identified as the wave function
for the collective motion, and
this identification will agree with the one that has been made in the
previous section, where
the description of large amplitude collective motion was
not tied to the BO approximation.  There we emphasized a connection
between the decoupled Hilbert space of the collective coordinates
and the crucial existence of a $K$-dimensional decoupled
coordinate manifold described by the functions $g^\alpha(Q,q=0)$.
The approximate separability of the associated
Hilbert space is based on the assumption that the functions
$[q|\nu\!:\!Q]$ describing the fast variables are confined
to a narrow region in $q$ space in the neighborhood of the collective
surface.  If this picture holds, we may expect the mathematical
details to work out reasonably.

Adopting the BO approximation, we set ourselves the task of finding
an effective Hamiltonian to describe the motion of the collective
variables.  This operator is defined, though not yet operationally,
by means of the equation,                   \begin{equation}
(n'|H_{{\rm eff}}(Q,P)|n)=\langle n'|\tilde{H}(\xi,\pi)|n\rangle.
\label{eq:2.6}                  \end{equation}
This relation will assume the status of a definition of the collective
Hamiltonian within the space of slow variables and $H_{{\rm eff}}$
recognized as a generalization of $H_C$ only after we
specify how to eliminate the fast variables from the right hand side.
The procedure that we shall follow is closely akin to the
traditional BO approach, with characteristic differences arising
from the facts that at the beginning we cannot specify which are the
slow and which the fast variables, and that the treatment of the
fast variables comes as a kind of afterthought, dependent in detail
on the prior treatment of the collective variables.

Let us start with the
potential energy,                     \begin{equation}
\tilde{V}(\xi)=\tilde{V}(g^\alpha(Q,q))\equiv V(Q,q),
\label{eq:2.7}                       \end{equation}
and evaluate the associated piece of (\ref{eq:2.6})  \begin{equation}
\langle n'|\tilde{V}(\xi)|n\rangle \equiv (n'|V_{eff}|n)  \\
\cong\int dQdq(n'|Q)[ 0\!:\!Q|q] \tilde{V}(
g^\alpha(Q,q))[q|0\!:\!Q]( Q|n),
\label{eq:2.8}                             \end{equation}
within the BO approximation.
The only feasible way, in general,
of integrating out the fast variables is to expand
$\tilde{V}$ in powers of $q^a$,               \begin{equation}
V(Q,q)=V(Q) +V_{,a}q^a+\half V_{,ab}q^a q^b +... \;,
\label{eq:2.9}                       \end{equation}
leading to                            \begin{equation}
V_{eff}(Q) = V(Q) + V^{(1)}(Q) + V^{(2)}(Q) +... \;,
\label{eq:2.10}                       \end{equation}
where assuming that the function $[q|0\!:\!Q]$ is normalized,
\begin{eqnarray}
V(Q)&=&V(Q,0),  \label{eq:2.11}    \\
V^{(1)}(Q)&=&V_{,a}\int|[q|0\!:\!Q] |^2 q^a \nonumber \\
& \equiv & V_{,a}(Q)\langle q^a\rangle_Q,  \label{eq:2.12} \\
V^{(2)}(Q) &=& \half V_{,ab}(Q)\langle q^aq^b\rangle_Q. \label{eq:2.13}
\end{eqnarray}

We thus see that the leading term is independent of the
wave function for the fast variables, coinciding with
the standard result for the potential energy of large amplitude
collective motion \cite{K1,KWDR} presented in the previous section.
The computation only requires the form of the collective
submanifold, $\xi^\alpha =g^\alpha(Q,0)$, which can be determined
by well-defined procedures \cite{KWDR}.  To go beyond
this lowest order, we need, besides the wave function of the
fast variables, to rearrange the expansion of V, as explained in
appendix B.  There it is shown that if we wish to interpret the small
quantities $q^a$ as components of a vector, we must replace the
ordinary second derivative by a covariant second derivative,
$V_{;ab}$, that can be computed from known or calculable quantities
according to the equations                 \begin{eqnarray}
V_{,a}(Q) &=& \tilde{V}_{,\alpha}g^\alpha_{,a}, \label{eq:2.14} \\
V_{;ab}(Q)&=&\tilde{V}_{;\alpha\beta}g^\alpha_{,a}g^\beta_{,b}
                       \label{eq:2.15} \end{eqnarray}

These equations are simplified by introducing one of the
decoupling conditions that follow from Eq.\ (\ref{eq:1.4}), namely that
(\ref{eq:2.14}) should be zero.  (Here we are
assuming that the two terms of (\ref{eq:1.4}) vanish separately.
Conditions for this to be true and modifications necessary when it
is not have been discussed in Ref.~\cite{KWDR})
The third term $V^{(2)}(Q)$, is the first, $\Delta^{(1)}V(Q)$,
of a sequence of contributions that we shall identify as
the leading corrections to the potential energy.  Further
discussion of the
evaluation of this term and of the additional correction terms, to be
identified below, will be continued later in this section, following
the
identification of all the pieces.

The remaining terms
will arise from the study of the kinetic energy, given after
transformation of coordinates by Eqs.\ (\ref{eq:2.19}) and
(\ref{eq:1.2}).
In order to integrate out the fast variables in the contribution that
these terms make to $H_{eff}$, consider the first term
of the kinetic energy.  We expand the mass
tensor in powers of $q$, and keep initially only the leading term
$B^{\mu\nu}(Q,0)\equiv B^{\mu\nu}(Q)$.
In this approximation, we first
restrict the study to the contribution of
those terms where the indices $\mu,\nu$ take on values $i,j$
in the collective subset.

As a preliminary to this calculation, we study the simpler object
\begin{eqnarray}
(n'|(P_i)_{{\rm eff}}|n)&=&\langle n'|P_i|n\rangle  \nonumber \\
&     =& (n'|(P_i -A_i)|n)      \nonumber \\
  & \equiv& (n'|D_i|n).    \label{eq:3.50}    \end{eqnarray}
Here $P_i$ is the collective momentum operator identified in the
previous section, and             \begin{equation}
A_i \equiv i\int dq [0\!:\!Q|q]\partial_i
[q|0\!:\!Q],
\label{eq:3.51}    \end{equation}
where $\partial_i$ means partial derivative with respect to $Q^i$.
Notice, however, that if we calculate straightforwardly,
\begin{eqnarray}
(n'|(P_iP_j)_{{\rm eff}}|n)&=&\langle n'|P_iP_j|n\rangle  \nonumber \\
&=& (n'|(P_iP_j -A_iP_j -A_jP_i + S_{ij})|n),  \label{eq:3.52}
                                    \end{eqnarray}
where                             \begin{equation}
S_{ij} = -\int dq [0\!:\!Q|q](\partial^2/\partial Q^i
\partial Q^j) [q|0\!:\!Q].  \label{eq:3.53} \end{equation}

As a consequence of this result, we are reassured that the effective
value of zero is zero, i.\ e.,            \begin{equation}
0=\langle n'|[P_i,P_j]|n\rangle =(n'|[P_i,P_j]|n).
\label{eq:3.54}                        \end{equation}
It is easy to see that the remaining canonical commutators also
project without change.

It is useful to rewrite (\ref{eq:3.52}) in a form that makes contact
with standard results.  We have    \begin{eqnarray}
(i\partial_j)[q|\nu\!:\!Q]i& =&\sum_{\nu'}
{}[q|\nu'\!:\!Q](A_j)(Q))_{\nu'\nu},   \label{eq:3.55}    \\
(A_j)_{\nu'\nu}&=&i\int\,dq[\nu'\!:\!Q|q]\partial_j[q|\nu\!:\!Q].
\label{eq:3.550}
                                      \end{eqnarray}
Because of the $Q$ dependence of the matrix element, it now follows
that
                                    \begin{eqnarray}
S_{ij}&=& \sum_\nu (A_i)_{0\nu}(A_j)_{\nu 0} +(i\partial_i)
A_j       \nonumber \\
&=& (i\partial_i)A_j +A_iA_j +S^{\prime}_{ij},
\label{eq:3.56}                  \end{eqnarray}        where
                            \begin{eqnarray}
(A_j)_{00} &=&A_j ,      \label{eq:3.57}   \\
S^{\prime}_{ij}&=& \sum_{\nu\neq 0}(A_i)_{0\nu}(A_j)_{\nu 0}.
                 \label{eq:3.58}       \end{eqnarray}
Consequently, we may also rewrite Eq.\ (\ref{eq:3.52}) as
                           \begin{equation}
(n'|(P_iP_j)_{{\rm eff}}|n) = (n'|(D_iD_j + S^{\prime}_{ij})|n).
              \label{eq:3.59}        \end{equation}
In the simplest case, where $[0\!:\!Q|q]$ is a real
wave function, and it follows that $A_i$ vanishes, the contribution
$S^{\prime}_{ij}$ remains to be taken into account.

We are now in a position to apply
to the computation of the collective kinetic energy
the same reasoning as just carried out for a product of momentum
operators.  Making use
of the analogue of (\ref{eq:3.59}), the result is
                           \begin{equation}
\frac{1}{8}\{D_i ,\{D_j ,B^{ij}(Q)\}\} +\half S^{\prime}_{ij}B^{ij}.
\label{eq:2.21}                                 \end{equation}
where the second term can be incorporated into the
collective potential energy as as a second such contribution,
$\Delta^{(2)}V(Q)$.

An additional contribution of this type is
obtained by setting $q^a=0$ in Eq.\ (\ref{eq:1.2}),  \begin{equation}
\Delta^{(3)}V(Q)= U(Q,0),   \label{eq:1.20}   \end{equation}
where $U(Q,q)$ is the quantum potential defined in Eqs.\
(\ref{eq:2.19})
and (\ref{eq:1.2}).

It remains for us to discuss the contributions from $T$ that depend
on $B^{ai}(Q)$ that ``mix" the collective and non-collective indices
and those that depend on the non-collective mass tensor $B^{ab}(Q)$.
The former can be neglected because of one of the decoupling conditions,
Eq.\ (\ref{eq:1.5}).
The leading contribution of the latter is seen
to be another contribution to the potential energy,  \begin{equation}
\Delta^{(4)} V =\half B^{ab}(Q)\langle p_a p_b\rangle_Q.
\label{eq:2.26}                               \end{equation}

To summarize our findings, we have derived the following effective
Hamiltonian,                               \begin{equation}
H_{{\rm eff}} = \frac{1}{8}\{D_i ,\{D_j ,B^{ij}(Q)\}\} + V(Q)
+\Delta V(Q),     \label{eq:2.27}         \end{equation}
where $\Delta V(Q)$ is the sum of four terms that summarize the leading
quantum corrections including the coupling to the fast variables,
\begin{equation}
\Delta V = \sum_{i=1}^4 \Delta^{(i)} V, \label{eq:2.28} \end{equation}
given respectively in or in relation to Eqs.\ (\ref{eq:2.13}),
(\ref{eq:2.21}), (\ref{eq:1.20}), and (\ref{eq:2.26}).
Let us contrast
this result with the corresponding form appropriate to the more
familiar molecular case.  The introduction of a curved
metric aside, the main difference is that in the molecular case, the
ground state wave function of the fast variables, for a fixed value of
$Q$, may be assumed known, and its eigenvalue, $\epsilon_0 (Q)$,
together with
$\Delta^{(2)} V$, contained in Eq.\ (\ref{eq:2.21}) constitutes the
collective  potential
energy \cite{B1,MSW}.

In the
present instance we cannot assume that we know the Hamiltonian of the
fast variables, except in an approximate sense that we now discuss.
We need to extend the considerations of the previous section
that led to the definition of the quantum
collective Hamiltonian operator $H_{C}$.  Instead
of specializing the transformed Hamiltonian operator to the
values $q^a=p_a=0$, we now retain terms up to second order in these
variables. In this treatment, we may replace $U(Q,q)$ by $U(Q,0)$, since
this term is already a small correction, and the remaining terms linear
in the fast variables may be dropped because of the decoupling
conditions.  To the specified accuracy, we obtain the following
quantum Hamiltonian,              \begin{eqnarray}
H&=&H_C +U(Q) +H_{NC},   \label{eq:2.50}   \\
H_{NC}&=&\half p_a p_b B^{ab}(Q) +\half q^a q^b V_{ab}(Q).
\label{eq:2.51}                 \end{eqnarray}

In appendix B, we point out that once the elements of $H_C$ have been
determined, one can choose a set of fast variables and
calculate the matrices that appear in (\ref{eq:2.51}).
For each value of $Q$, the Hamiltonian $H_{NC}$ represents a standard
normal mode problem.  Let $c_\alpha,c^{\dag}_\alpha$ be normal mode
destruction and creation operators, $\Omega_\alpha$ the corresponding
frequencies, and $\hat{n}_\alpha=c^{\dag}_\alpha c_\alpha$.  Assuming
local stability, i.\ e., $\Omega_\alpha$ real and positive, we have
                                        \begin{equation}
H_{NC}(Q)=\sum_\alpha (\hat{n}_\alpha +\half)\Omega_\alpha(Q).
\label{eq:2.60}                       \end{equation}
The practical importance of the contribution of (\ref{eq:2.60}) has
been noted in several applications carried out in the
past \cite{WKDC,UK}.

The quantum Hamiltonian, expressed in terms of the optimum choice of
variables, but in a restricted approximation, has thus been found.
Except for the term $U(Q)$, it has been expressed in terms of elements
that can be calculated.  Evaluation of $U(Q)$ appears to require some
properties of the point transformation that have not so far been
studied.  Comparison of approximate calculations (with this term
omitted)
with exact calculations indicates that it is probably a small
correction.
As previously remarked, it is of order $N^{-2}$  compared to the
main term
in the potential energy, where $N$ is the number of degrees of
freedom.         We shall make no further allusion to this term
in the present paper.

Let us now return to a discussion of the correction terms, $\Delta V$
by which $H_{{\rm eff}}$, Eq.\ (\ref{eq:2.27}), differs from $H_C$.
We have just discarded
$\Delta^{(3)} V$. The sum $\Delta^{(1)} V + \Delta^{(4)} V$ has been
seen in (\ref{eq:2.60}) to be a sum of oscillator terms in the
approximation considered, its contribution then depending, naturally,
on the state of motion of the fast variables. As stated above, in
this section we shall consider only the lowest energy state for these
variables, so that the contribution of this term is just the zero-point
energy.

It remains for us to discuss the contribution of the term
$\Delta^{(2)} V$, the Berry scalar potential, associated with the
projection of the kinetic
energy onto the collective subspace.  This contribution goes together
with that from
the ``vector potential", $A_i$.  In Sec.\ \ref{sec:models},
we shall study
examples where these terms contribute.  In fact, the manner in which
 both
the vector and scalar potentials contribute is best studied within the
context of these illustrations.

In this section we have thus gone as far as we shall in defining a
quantum
theory of the slow variables, having derived an effective Hamiltonian
and shown that it is expressed in terms of canonical variables.  Thus
we are also free to study this problem with the help of the Heisenberg
equations of motion.  In the next section we shall consider variational
formulations of these equations.

\setcounter{equation}{0}
\section{Variational principles; Lagrangian formulation
\label{sec:varia}}

Having defined an effective quantum theory within the collective
subspace, we can imagine that we study this theory by means of
Heisenberg's equations of motion,               \begin{eqnarray}
-i[Q^i ,H_{{\rm eff}}] &=& \frac{\partial}{\partial P_i}
H_{{\rm eff}}  ,
\label{eq:4.1}\\
i[P_i ,H_{{\rm eff}}] &=& \frac{\partial}{\partial Q^i}
H_{{\rm eff}}   .
\label{eq:4.2}                        \end{eqnarray}
The purpose of this section is to show that these equations can be
derived from a variational principle, the so-called trace variational
principle, involving $H_{{\rm eff}}$, or an operator closely related
to it;
it is to show, further,
that a reworking of this principle leads to a second version of the
variational principle, involving an effective Lagrangian,
$L_{{\rm eff}}$, that is defined in the natural way as the Legendre
transform of $H_{{\rm eff}}$.  According to Moody, Schapere, and
Wilzek \cite{MSW}, when $L_{{\rm eff}}$ is defined by means of a
path integral,
it is not exactly the Legendre transform of $H_{{\rm eff}}$.  We
shall not
encounter any such difficulty as long as the mass tensor is independent
of coordinates, the only case studied by previous authors.
When the mass tensor
enters in full generality,
the formulation of a variational principle of the type we have in mind
is not quite as straightforward as in previous instances that we have
studied \cite{KLV,LK2}.

In order not to have to deal with all subtleties at once, let us first
consider the case that the mass tensor does not depend on coordinates.
In that case we may utilize the standard version    \begin{equation}
\delta {\rm Tr}\{H_{{\rm eff}} -i\Lambda[Q^i ,P_i]\}=0.
\label{eq:4.3}                                     \end{equation}
Here the trace is taken over a {\em finite} subspace of the collective
Hilbert space.  This assumption allows us to utilize the invariance
of the trace under cyclic permutation, a property that plays an
essential role in the manipulation of the second term of
(\ref{eq:4.3}).
In this term, $\Lambda$ is a Lagrange multiplier operator associated
with a set of constraints that are the non-vanishing canonical
commutation relations.  These, at least, must be imposed because, to
start with, the variational quantities are arbitrary matrix elements
within the space included in the trace, and these variations must be
subject to the kinematical constraints.  The fact that we can limit
the constraints to the commutators for canonical pairs  is, at the
moment, an observation.

In order to come out with the Heisenberg equations of motion upon
variation, it is necessary, as well, to choose variations of the
coordinates that depend only on the coordinates themselves, so
as to validate the formula                        \begin{equation}
\delta V(Q) = (\partial V(Q)/\partial Q^k)\delta Q^k.
\label{eq:4.4}                                 \end{equation}
The equations of motion that follow from these assumptions are
\begin{eqnarray}
-i[Q^i ,\Lambda] &=& \frac{\partial H_{{\rm eff}}}{\partial P_i},
\label{eq:4.5}\\
i[P_i ,\Lambda] &=& \frac{\partial H_{{\rm eff}}}{\partial Q^i}.
\label{eq:4.6}                      \end{eqnarray}
To make these agree with Heisenberg's equations, the choice
$\Lambda=H_{{\rm eff}}$ is indicated.

We apply this formalism to the case with non-vanishing Berry
potentials but constant mass matrix, described by the Hamiltonian
operator,                        \begin{equation}
H_{{\rm eff}} =\half D_iD_jB^{ij} + V,  \label{eq:4.50}
\end{equation}
where it is understood that $V$ may include the correction terms
discussed in the previous section.  The operator equations of motion
that follow from (\ref{eq:4.5}), (\ref{eq:4.6}), and (\ref{eq:4.50}),
after some standard manipulations, are  \begin{eqnarray}
\dot{Q}^i &=& D_jB^{ji},  \label{eq:4.51}   \\
\dot{D}_i &=& -V_{,i}+ \half\{\dot{Q}^j,{\cal F}_{ij}\},
\label{eq:4.52}
       \end{eqnarray}
where
\begin{eqnarray}
\dot{A}_i &=& \half \{(\partial A_i/\partial Q^j),\dot{Q}^j\},
\label{eq:4.53} \\
{\cal F}_{ij} &=& (\partial A_i/\partial Q^j) -(\partial A_j/
\partial Q^i),      \label{eq:4.54}       \end{eqnarray}
and                                    \begin{equation}
D_i = \dot{Q}^jB_{ji},     \label{eq:4.55}  \end{equation}
with $B_{ji}$ the matrix inverse to $B^{ji}$, is the solution of
Eq.\ (\ref{eq:4.51}) for the gauge-covariant momentum.

 From the combination of (\ref{eq:4.52}) and (\ref{eq:4.55}), we can
derive the familiar Lagrangian equations of motion
\begin{equation}
\ddot{Q}^jB_{ji} +V_{,i} -\half\{{\cal F}_{ji},\dot{Q}^j\} = 0.
\label{eq:4.56}                 \end{equation}
Of some interest is that the associated variational principle from
which these equations can be derived, namely   \begin{eqnarray}
0 &=& \delta{\rm Tr}(-L_{{\rm eff}})   \nonumber \\
  &=& \delta{\rm Tr}(-\half\dot{Q}^i\dot{Q}^jB_{ji} -\dot{Q}^iA_i +V),
\label{eq:4.57}                \end{eqnarray}
can be obtained
directly from the starting trace variational principle,
(\ref{eq:4.3}), simply by eliminating the momenta in favor of the
velocities by means of the relations given above.  Instead of
independent
variations of coordinates and momenta, only the coordinates are to be
varied. In this variation, the velocity $\dot{Q}^i$ is to be replaced
by
$-i[Q^i,H_{{\rm eff}}]$ and $H_{{\rm eff}}$ is not to be varied.
               One obtains the
correct equations of motion because algebraic manipulation of the
resulting trace expressions, containing commutators with
$H_{{\rm eff}}$,
that is aimed at isolating the coordinate
variations, produces the same results as integration by parts of the
time derivatives in the corresponding classical variational principle,
proper attention being paid to the order of non-commuting factors.
For the case just concluded, the outcome is thus completely
satisfactory.

When we reinstate the $Q$ dependence of the mass tensor, an
additional problem is encountered.  If we insist on the cyclic
invariance of the trace, which has played an essential role in the
previous applications of the trace variational principle, then we have
\begin{equation}
{\rm Tr}\{P_i ,\{P_j , B^{ij}\}\} = 2{\rm Tr}\{P_i P_j, B^{ij}\}.
\label{eq:4.7}                          \end{equation}
However, direct transformation, using the commutation relations,
yields
\begin{equation}
\{P_i ,\{P_j , B^{ij}\}\} = 2\{P_i P_j, B^{ij}\} +B^{ij}_{,ij},
\label{eq:4.8}                          \end{equation}
leading to an apparent contradiction upon formation of the trace.
Of course, in deriving (\ref{eq:4.8}), we have used the commutation
relations appropriate to an infinite Hilbert space.  The point is the
same as that ${\rm Tr}QP\neq {\rm Tr}PQ$ for a truly canonical pair.
In order to derive
the equations of motion from a trace variational principle, we deal
with this problem by adopting
(\ref{eq:4.7}), but including the last term in (\ref{eq:4.8})
as part of the potential energy. This implies the assumption that the
trace is taken, to start with, only over a finite-dimensional vector
space and that therefore the invariance of the trace under cyclic
permutation is a correct operation.  As a consequence, the equations
of motion are given in matrix form. The operator form of the
equations of motion is recognized as a limit of these equations.
In this limit, the application of (\ref{eq:4.8})
{\em after variation}, can be seen to yield the equations
of motion in the desired symmetrical (Weyl) structure.
Thus the final variational principle takes the form
\begin{equation}
\delta {\rm Tr} \{H_{{\rm mod}}-iH_{{\rm eff}}[Q^i ,P_i ]\},
\label{eq:4.9}
\end{equation}                where          \begin{equation}
H_{{\rm mod}} = H_{{\rm eff}} + \frac{1}{8}B^{ij}_{,ij}.
\label{eq:4.10}
\end{equation}

In the discussion above, we have noted a problem only in connection
with the part of the kinetic energy quadratic in the momenta.  It may
be verified that a similar difficulty does not occur for the terms
that are bilinear in the momenta and the ``vector potential".

Unfortunately, in the transition from Hamilton's equations to
Lagrange's
equations for this case additional complications are encountered;
because of the
non-commutativity of the mass tensor with the momentum, the Lagrange
equations do not follow from the expression obtained
by eliminating the momenta in favor of the velocities in the starting
variational principle.  The equations of motion that follow from this
conjectured Lagrangian variational expression differ from the correct
equations of motion by additional ``quantum" potentials.  This line
of inquiry does not appear to be illuminating and therefore will not
be pursued.

\setcounter{equation}{0}
\section{Illustrative models \label{sec:models}}

It is simplest to illustrate the main points by choosing models in
which the collective coordinates have already been identified, so that
we need not enter into the intricacies of the theory of large amplitude
collective motion {\em per se}.  For instance, in the first model
studied
below, there are no terms in the Hamiltonian linear in the fast
variables.
Thus the model satisfies the decoupling conditions exactly.
In both of the
models to be studied, we break time-reversal invariance,
and in that sense
they are somewhat artificial.

\subsection{The first model: Berry phase in excited states}

We study the Hamiltonian
\begin{eqnarray}
H&=& H_{{\rm core}} + H_{{\rm sp}} + H_{{\rm int}}  \nonumber \\    &=&
H_{{\rm core}} + H_{{\rm NC}} \label{eq:5.1}  \\
H_{{\rm core}}&=& \frac{{\bf P}^2}{2M} + V(Q),  \label{eq:5.2} \\
H_{{\rm sp}}&=& \omega(Q)(a_1^{\dag}a_1 +a_2^{\dag}a_2 +1),
\label{eq:5.3}\\
H_{{\rm int}}&=& -G(Q_- a_1^{\dag}a_2 +Q_+ a_2^{\dag}a_1).
\label{eq:5.4}
\end{eqnarray}
Here the coordinates ${\bf Q}$ and the canonical momenta ${\bf P}$
are both two-dimensional
vectors.  We use the notation ${\bf Q}=(Q_1,Q_2)$,
$Q_{\pm}=Q_1\pm iQ_2$, and
$Q=(Q_1^2 +Q_2^2)^{1/2}$.  (In this section, we are not
maintaining the
distinction between upper and lower indices.)
Furthermore, the $a_i,a_i^{\dag},i=1,2$
are boson destruction and creation operators, $G$ is a coupling
strength,
and the frequency, $\omega(Q)$, of the uncoupled boson modes has
been given
a so far unspecified dependence on $Q$ that will be chosen for
analytic and
numerical convenience.  Note that the
Hamiltonian, (\ref{eq:5.1}) conserves the boson number,
\begin{equation}
N = a_1^{\dag}a_1+ a_2^{\dag}a_2 = {\rm constant}.   \label{eq:5.5}
\end{equation}

Since the $N=0$
problem is completely trivial,  the first interesting case for the
present
model is $N=1$.
Here, the state vectors may be written exactly as a superposition
\begin{equation}
|n\rangle = \int d{\bf Q}\{({\bf Q}|n1)a_1^{\dag}|0]
+ ({\bf Q}|n2)a_2^{\dag}|0]\},    \label{eq:5.6}
\end{equation}
where $|0]$ is the vacuum state.  The use of square brackets for the
 vacuum
state of the fast variables is consistent with the notation introduced
in
Sec.\ \ref{sec:BO}. The resulting eigenvalue
equation in the space of the collective variables, ${\bf Q}$, is
determined
by the two by two effective Hamiltonian, $H_{eff}$, with matrix
elements
\begin{eqnarray}
(H_{eff})_{11}&=&H_{core} +2\omega(Q)=(H_{eff})_{22}, \nonumber \\
(H_{eff})_{12}&=&-GQ_-=(H_{eff})_{21}^{\ast}. \label{eq:5.7}
\end{eqnarray}
Below we shall describe our solutions of the associated Schr\"{o}dinger
equation.

These exact solutions are to be compared with the adiabatic
approximation,
with and without the Berry potential terms.  For this approximation, we
require the normal modes of $H_{NC}$, which we calculate in a
standard way
from the equations of motion,          \begin{eqnarray}
[a_1,H_{NC}]&=& \omega a_1 - GQ_- a_2,    \nonumber \\
{}[a_2,H_{NC}]&=& \omega a_2 - GQ_+ a_1,   \label{eq:5.8}
\end{eqnarray}
by forming the matrix elements,       \begin{equation}
\psi_i = [0|a_i|\Psi].   \label{eq:5.9}   \end{equation}
With $\Omega$ representing the energy of the state $|\Psi]$,
we obtain the equations         \begin{eqnarray}
\Omega \psi_1 &=& \omega \psi_1 -GQ_-\psi_2, \nonumber \\
\Omega \psi_2 &=& -GQ_+\psi_1 +\omega\psi_2, \label{eq:5.10}
\end{eqnarray}
that yield the eigenvalues        \begin{eqnarray}
\Omega^{(1)}(Q) &=& \omega(Q) - GQ,  \nonumber \\
\Omega^{(2)}(Q) &=& \omega(Q) + GQ,   \label{eq:5.11}  \end{eqnarray}
that are degenerate when $Q=0$.
The associated normalized solutions of (\ref{eq:5.10}) are
represented most
conveniently by introducing the normal-mode creation operators
\begin{equation}
b_i^{\dag} = \psi^{(i)}_j a_j^{\dag}.    \label{eq:5.12}
\end{equation}
In detail we have as a possible choice,
\begin{eqnarray}
b_1^{\dag} &=& \frac{1}{\sqrt{2}}a_1^{\dag}+
\frac{1}{\sqrt{2}}\exp{i\phi({\bf Q})}a_2^{\dag}, \label{eq:5.13} \\
b_2^{\dag} &=& \frac{1}{\sqrt{2}}\exp{i\phi({\bf Q})}a_1^{\dag} -
\frac{1}{\sqrt{2}}a_2^{\dag}, \label{eq:5.14}    \end{eqnarray}
where                  \begin{equation}
\tan\phi({\bf Q})=Q_2/Q_1.   \label{eq:5.15} \end{equation}

We now apply these elementary results to the adiabatic approximation.
In this case we represent a suitable subset of the eigenfunctions
(\ref{eq:5.6}) in the form                  \begin{equation}
|n\rangle = \int d{\bf Q} ({\bf Q}|n)b_1^{\dag}|0\!:\!Q],
\label{eq:5.16}
\end{equation}
where the notation $0\!:\!Q$ refers to the vacuum for the normal modes.
(For the current model, it coincides with the uncoupled vacuum.)
The considerations of Sec.\ \ref{sec:BO} now apply to this
class of state vectors and, in particular, we apply
Eq.\ (\ref{eq:3.59}).  As a special case of this equation, we have
\begin{equation}
\sum_i (P_i)^2 \rightarrow \sum_i (P_i -A_i)^2 +\sum_i |(A_i)_{21}|^2,
\label{eq:5.160}
\end{equation}
where
\begin{eqnarray}
A_i &=& i[0\!:\!Q|b_1\partial_i
b_1^{\dag}|0\!:\!Q],  \label{eq:5.17}   \\
(A_i)_{21} &=& i[0\!:\!Q|b_2\partial_i
b_1^{\dag}|0\!:\!Q],     \label{eq:5.18}   \end{eqnarray}
where $\partial_i$ means the partial derivative with respect to $Q_i$.
With the help of Eqs.\ (\ref{eq:5.13}) and (\ref{eq:5.14}), the
quantities
of interest are found to take the values       \begin{eqnarray}
{\bf A} &=& \frac{1}{2Q^2}(Q_2,-Q_1),  \label{eq:5.19} \\
\sum_i|(A_i)_{21}|^2 &=& (1/4Q^2).  \label{eq:5.20}
\end{eqnarray}
The singular character of these results was to be expected.

The collective or adiabatic Hamiltonian, $H_C$, that thus emerges from
the assumption that the state vectors of interest can be written in the
form (\ref{eq:5.16}), has the structure
\begin{equation}
H_C = (1/2M)[({\bf P}-{\bf A})^2 +({\bf A}_{21})^2] +V(Q) )
+(3/2)\Omega^{(1)}(Q)+(1/2)\Omega^{(2)}(Q). \label{eq:5.21}
\end{equation}

We turn to the problem of solving the associated eigenvalue problem.
We wish to compare the results of exact (numerical) calculations
with the eigenvalues of the collective Hamiltonian. It is useful
to choose $\omega(Q)$ such that the latter is exactly solvable.
One such choice is
\begin{equation}
\omega(Q) = \frac{1}{2} GQ .
\end{equation}
In this case the additional contribution to the potential
is zero,
\begin{equation}
\frac{3}{2}\Omega^{(1)}(Q)+\frac{1}{2} \Omega^{(2)}(Q)
= 0.
\end{equation}
We further take $M=1$ and $V(Q)=\frac{1}{2}Q^2$.

We use the definition of the two-dimensional angular momentum,
\begin{eqnarray}
L& =&- i \frac{\partial}{\partial \phi}
\nonumber \\ &=&
-i (Q_1 P_2-Q_2P_1),
\end{eqnarray}
to simplify the ${\bf P}\cdot{\bf A}$ term.
We can further simplify the equation
$H_c \psi(\vec Q) = E \psi(\vec Q)$
by substituting
\begin{equation}
\psi(\vec Q) = Q^{-1/2} \chi(Q) {\rm e}^{im\phi}
\end{equation}
and find
\begin{equation}
\left\{
-\frac{1}{2} \frac{d^2}{d Q^2}
+ \frac{1}{2Q^2}[m^2+m+\frac{1}{4}]
+\frac{1}{2} Q^2\right\} \chi(Q) =
E\chi(Q)
\end{equation}
Even though the centrifugal term has changed, this is still
very similar to
 the  radial equation for the two-dimensional harmonic
oscillator. It can be solved by the substitution
\begin{equation}
\chi(q) = q^{\alpha+1/2} {\rm e}^{-q^2/2} L_n^{(\alpha)}(q^2).
\end{equation}
As is well known the right hand side of this equation satisfies the
condition
\begin{equation}
- \frac{d^2}{d q^2} \chi
+(q^2+\frac{1-4\alpha^2}{4q^2}) \chi = (4n+2\alpha+2) \chi.
\end{equation}
We thus find that
\begin{equation}
\alpha(m) = \sqrt{1/4+(m+1/2)^2}
\end{equation}
and
\begin{equation}
 E_{nm} = (2n+\alpha(m)+1).
\label{eq:EnmB}
\end{equation}
Without the Berry's phase terms we would have found
\begin{equation}
 E_{nm} = (2n+|m|+1).
\label{eq:EnmnoB}
\end{equation}
When $G=0$ the solution (\ref{eq:EnmnoB}) is exact.
Thus Eq.~(\ref{eq:EnmB}), which is independent
of $G$, cannot be valid for all $G$. It should be
valid in the adiabatic limit, which means that the two
frequencies $\Omega$ must be very different. This
occurs when $G$ is very large.

The exact solution of the problem can be calculated using
a spherical harmonic oscillator basis for the $Q$ coordinates.
This is coupled to a state containing either one $a^\dagger_1$ or
one $a^\dagger_2$ boson,
\begin{equation}
|n,m,n_1,n_2\rangle = |n,m) (a^\dagger_1)^{(n_1)}
(a^\dagger_2)^{(n_2)}|0].
\end{equation}
The interaction Hamiltonian only couples states with $n_1=1,m=m_1$
to states with $n_2=1,m=m_1+1$. If we thus choose the value $m_1$
we have only a relatively small matrix to diagonalize. Since we
wish to obtain accurate results for large $G$, we allow the number of
$Q$
harmonic oscillator quanta to be fairly large. We have used
up to 200 harmonic oscillator states in each block (which leads
to a 400 by 400 matrix eigenvalue problem).

In table \ref{tab:model1.1} we give a selected set of results
for $m_1=0$ and a number of values of $G$.
Similar results for $m_1$=1 are listed in table \ref{tab:model1.2}.
We clearly see the
convergence to the the collective model results including the
Berry phase for large $G$.

\begin{table}[htb]
\caption{The lowest 10 eigenvalues of the coupled problem for
$m_1=0$ as a function of $G$. The column labeled Berry lists the
eigen energies of the collective Hamiltonian \protect{\ref{eq:5.21}}.
\label{tab:model1.1}}
\begin{center}
\begin{tabular}{rrrrrrrr}
\hline
G &
0.01     & 0.1      & 1           & 10        & 100       & 1000
&Berry\\
\hline
$E_1$ &
1.00875  & 1.07849  &   1.36229   &   1.58463 &   1.66532 &   1.69302
& 1.7071\\
$E_2$ &
2.01328  & 2.13134  &   2.92007   &   3.48844 &   3.63505 &   3.68298
& 3.7071\\
$E_3$ &
3.01540  & 3.14492  &   4.09426   &   5.39473 &   5.60870 &   5.67438
& 5.7071\\
$E_4$ &
4.01827  & 4.18125  &   5.30157   &   7.29413 &   7.58443 &   7.66658
& 7.7071\\
$E_5$ &
5.01997  & 5.19061  &   6.57000   &   9.17666 &   9.56145 &   9.65932
& 9.7071\\
$E_6$ &
6.02221  & 6.22068  &   7.66964   &   11.0262 &   11.5393 &   11.6525
& 11.7071\\
$E_7$ &
7.02367  & 7.22768  &   8.90918   &   12.8133 &   13.5177 &   13.6459
& 13.7071\\
$E_8$ &
8.02558  & 8.25425  &   10.0230   &   14.4981 &   15.4965 &   15.6396
& 15.7071\\
$E_9$ &
9.02687  & 9.25972  &   11.1883   &   16.0926 &   17.4754 &   17.6335
& 17.7071\\
$E_{10}$ &
10.0286  & 10.2840  &   12.3410   &   17.7119 &   19.4544 &   19.6276
& 19.7071\\
\hline
\end{tabular}
\end{center}
\end{table}

\begin{table}[htb]
\caption{The lowest 10 eigenvalues of the coupled problem for
$m_1=1$ as a function of $G$. The column labeled Berry lists the
eigen energies of the collective Hamiltonian \protect{\ref{eq:5.21}}.
\label{tab:model1.2}}
\begin{center}
\begin{tabular}{rrrrrrrr}
\hline
G &
0.01     & 0.1      & 1           & 10        & 100       & 1000
&Berry\\
\hline
$E_1$ &
   2.01308 &   2.11333&   2.41566&   2.54921&   2.57601&   2.58042
& 2.58114\\
$E_2$ &
   3.01670 &   3.17369&   4.20360&   4.51926&   4.57057 &   4.57959
& 4.58114\\
$E_3$ &
   4.01807 &   4.16340&   5.44213&   6.48766&   6.56479&   6.57866
& 6.58114\\
$E_4$ &
   5.02086 &   5.21512&   6.43129&   8.45380&   8.55880&   8.57766
& 8.58114\\
$E_5$ &
   6.02202 &   6.20307&   7.86356&   10.4165&   10.5527&   10.5766
& 10.58114\\
$E_6$ &
   7.02436 &   7.25000&   8.79958&   12.3739&   12.5464&   12.5755
& 12.58114\\
$E_7$ &
   8.02538 &   8.23691&   10.1070&   14.3218&   14.5400&   14.5744
& 14.58114\\
$E_8$ &
   9.02745 &   9.28063&   11.1975&   16.2512&   16.5336&   16.5733
& 16.58114\\
$E_9$ &
   10.0284 &   10.2669&   12.3141&   18.1346&   18.5270&   18.5721
& 18.58114\\
$E_{10}$ &
   11.0302 &   11.3082&   13.5364&   19.8627&   20.5204&   20.5709
& 20.58114\\
\hline
\end{tabular}
\end{center}
\end{table}

\subsection{The second model: ground state}

We next study a model that is also exactly decoupled, differing
from the one just investigated only in
the form of the interaction.  In this model Eqs.\
(\ref{eq:5.1})-(\ref{eq:5.3})
stand as given, but Eq.\ (\ref{eq:5.4}) is modified to
\begin{eqnarray}
H_{int} &=& -G_0 Q(a_1^{\dag}a_2 + a_2^{\dag}a_1)
 -G_1(Q_+ a_1 a_2 +Q_- a_1^{\dag}a_2^{\dag}).  \label{eq:5.22}
\end{eqnarray}
The second term of this interaction
spoils the conservation of boson number used to simplify the
solution of the
previous problem.  For the present problem, it is already
interesting to study
the spectrum when the fast variables are in their ground levels,
since there
 are now non-trivial ground-state correlations and associated
non-trivial
Berry potentials, that we shall calculate.

We first turn to the study of the BO approximation.  As before we
need  the normal modes of the fast variables, as determined from the
the equations of motion
\begin{eqnarray}
[a_1,H_{NC}] &=& \omega a_1 -G_0Q a_2 -G_1Q_- a_2^{\dag}, \nonumber \\
{}[a_2,H_{NC}] &=& \omega a_2 -G_0Q a_1 -G_1Q_- a_1^{\dag},
\label{eq:5.24}
\end{eqnarray}
and their hermitian conjugates.  In terms of the definitions
\begin{eqnarray} b_i^{\dag}&=&\psi_j^{(i)}a_j^{\dag}-\chi_j^{(i)}a_j,
\label{eq:5.25} \\ \psi_j^{(i)}&=&[0\!:\!Q|a_j|\Psi^{(i)}],
\label{eq:5.26}\\ \chi_j^{(i)}&=&[0\!:\!Q|a_j^{\dag}|\Psi^{(i)}],
\label{eq:5.27}\\ |\Psi^{(i)}] &=&b_i^{\dag}|0\!:\!Q], \label{eq:5.28}
\end{eqnarray}
we obtain the eigenvalue equations \begin{eqnarray}
\Omega\psi_1&=&\omega\psi_1 -G_0Q\psi_2 -G_1Q_-\chi_2, \nonumber \\
\Omega\psi_2&=&\omega\psi_2 -G_0Q\psi_1 -G_1Q_-\chi_1, \nonumber \\
-\Omega\chi_1&=&\omega\chi_1 -G_0Q\chi_2 -G_1Q_+\psi_1,\nonumber \\
-\Omega\chi_2&=&\omega\chi_2 -G_0Q\chi_1 -G_1Q_+\psi_1.\label{eq:5.29}
\end{eqnarray}
The physical eigenvalues are the positive roots of
\begin{equation}
\Omega^2(Q) = (\omega \mp G_0Q)^2 -G_1^2Q^2, \label{eq:5.30}
\end{equation}
which are again degenerate at $Q=0$.
As for the first model studied,
we label these solutions $\Omega^{(i)},i=1,2$.
The corresponding amplitudes are determined from the equations of
motion and
the normalization conditions
\begin{equation}
|\psi_1|^2 +|\psi_2|^2 -|\chi_1|^2 -|\chi_2|^2 =1. \label{eq:5.31}
\end{equation}
The simplest forms for the amplitudes are achieved by
repeated use of the eigenvalue equation (\ref{eq:5.30}).  We thus find
\begin{eqnarray}
\psi_2^{(j)}&=&(-1)^{j+1}\psi_1^{(j)}={\rm real}, \label{eq:5.32} \\
\chi_i^{(j)}&=&\exp(i\phi)\tilde{\chi}_i^{(j)}, \label{eq:5.33} \\
\tilde{\chi}_2^{(j)}&=&(-1)^{j+1}\tilde{\chi}_1^{(j)}={\rm real},
\label{eq:5.34} \\
\psi_1^{(j)}&=&
\frac{(\Omega^{(j)}+\omega)+(-1)^jG_0Q}
{\sqrt{2}[(\omega+\Omega^{(j)})^2+(-1)^j2G_0Q(\omega+\Omega^{(j)})
+(G_0^2-G_1^2)Q^2]^{1/2}},
\label{eq:5.35} \\
\chi_1^{(j)}&=&
 (-1)^{j+1}f^{(j)} (Q)\psi_1^{(j)},
\label{eq:5.36}
\\
f^{(j)}&=&\frac{G_1Q}{(\Omega^{(j)}+\omega)+(-1)^jG_0Q}.
\end{eqnarray}
Here $\phi$ is the polar angle for $Q$, $\phi = \arctan (Q_2/Q_1)$.

These results are thus available for application to the BO
approximation studied by means of the assumption
\begin{equation}
|n\rangle =\int d{\bf Q}|{\bf Q}\rangle ({\bf Q}|n)|0\!:\!Q].
\label{eq:5.39}
\end{equation}
It follows that the Berry vector potential is given by the formula
\begin{equation}
A_i = i[0\!:\!Q|\partial_i|0\!:\!Q].
\label{eq:5.40}
\end{equation}
To carry out this calculation, we need the form of the correlated
vacuum state,
as given by the equation \cite{Sa}
\begin{equation}
|0\!:\!Q] = {\cal N} \exp[\half Z_{ij}a_i^{\dag}a_j^{\dag}]|0],
\label{eq:5.41}
\end{equation}
where ${\cal N}$ is a real normalization factor, whose value we
shall not need,
$a_i |0]=0$, and $Z_{ij}$ is a generally complex array whose
values will
be considered below.  Using the constancy of the norm of the
correlated ground
state as a function of the values of ${\bf Q}$, we can rewrite
(\ref{eq:5.40})
as
\begin{equation}
A_i =\half i \{[ 0\!:\!Q|\partial_i|0\!:\!Q]
- [\partial_i 0\!:\!Q|0\!:\!Q]\}, \label{eq:5.42}
\end{equation}
which shows that the (real) normalization factor ${\cal N}$ does not
contribute. This leaves the result
\begin{equation}
A_i = \frac{1}{4}i[\langle a_k^{\dag}a_l^{\dag}\rangle\partial_i
Z_{kl} - \langle a_k a_l\rangle\partial_i Z_{kl}^{\ast}],
\label{eq:5.43}
\end{equation}
where the averages, indicated by angular brackets,
are with respect to the correlated vacuum.

We next indicate how (\ref{eq:5.43}) can be evaluated in terms of the
solutions
we have found above for the equations of motion.  The quantities
$Z_{kl}$
are solutions of the equations
\begin{equation}
\psi_j^{(i)}Z_{jk}^{\ast} = \chi_k^{(i)},    \label{eq:5.44}
\end{equation}
that follow from the definition $b_i |0\!:\!Q] =0.$  The expectation
values
can also be evaluated with the help of the formulas
\begin{equation}
a_i = \psi_i^{(j)}b_j +\chi_i^{(j)\ast}b_j^{\dag},\label{eq:5.45}
\end{equation}
and their hermitian conjugates, that are the inverse to
Eq.\ (\ref{eq:5.25}).
  An elementary calculation now yields
\begin{equation}
\langle a_k^{\dag} a_l^{\dag}\rangle = \langle a_k a_l\rangle^{\ast}
= \chi_k^{(j)}\psi_l^{(j)\ast}.     \label{eq:5.46}
\end{equation}
Thus for the vacuum Berry potential, we obtain the formula
\begin{equation}
A_i = \frac{1}{4}i[\chi_k^{(m)}\psi_l^{(m)\ast}\partial_i
Z_{kl} -\chi_k^{(m)\ast}\psi_l^{(m)}\partial_i Z_{kl}^{\ast}].
\label{eq:5.47}
\end{equation}

Any other matrix element, $(A_i)_{\nu 0}$ of the Berry potential can be
calculated by similar techniques.  In fact, the only non-vanishing
elements
of this type are $(A_i)_{11,0}$, $(A_i)_{20,0}$, and $(A_i)_{02,0}$,
where,
for example, $\nu=20$ means a state with two correlated bosons of type 1
and none of type 2.  We quote formulas for these matrix elements that
follow
from the same elementary techniques that furnished Eq.\ (\ref{eq:5.47}).
We thus have
\begin{eqnarray}
(A_i)_{11,0}&=&i[11\!:\!Q|\partial_i |0\!:\!Q] \nonumber \\
&=& -i[\partial_i\; 11\!:\!Q|0\!:\!Q] \nonumber \\
&=& -i[0\!:\!Q|b_1\partial_i(b_2 )|0\!:\!Q]\} \nonumber \\
&=&-i
[\sum_{j}\psi_j^{(1)\ast}\partial_i\chi_j^{(2)\ast}
-\chi_j^{(1)\ast}\partial_i\psi_j^{(2)\ast}].
\label{eq:5.48}   \end{eqnarray}
Similarly, remembering the normalization of the states, we find
\begin{equation}
(A_i)_{20,0}= \frac{1}{i\sqrt{2}}[\sum_{j}
\psi_j^{(1)\ast}\partial_i\chi_j^{(1)\ast}
-\chi_j^{(1)\ast}\partial_i\psi_j^{(1)\ast}].
\label{eq:5.490}      \end{equation}
Finally, the matrix element $(A_i)_{02,0}$ is obtained from

(\ref{eq:5.490})
simply by replacing all superscripts 1 by 2.
In order to implement fully the BO approximation, it is necessary to
evaluate these matrix elements in terms of the explicit solutions of
the
model displayed in Eqs.\ (\ref{eq:5.32})-(\ref{eq:5.35}).  It is also
appropriate to remark here that in consequence of these relations and
Eq.\ (\ref{eq:5.44}) we can write
\begin{equation}
Z_{jk} = \exp(-i\phi)\tilde{Z}_{jk},\;\;\tilde{Z}_{jk}={\rm real}
\label{eq:5.491}
\end {equation}

The calculation of the Berry potentials as defined above is now
straightforward.  In terms of purely real quantities, we find,
for example,
\begin{equation}
A_i = (\partial_i \phi)[(\tilde{\chi}_1^{(1)}\psi_1^{(1)}+
\tilde{\chi}_1^{(2)}
\psi_1^{(2)})\tilde{Z}_{11} + (\tilde{\chi}_1^{(1)}\psi_1^{(1)}
-\tilde{\chi}_1^{(2)}\psi_1^{(2)})\tilde{Z}_{12}].  \label{eq:5.492}
\end{equation}     Using the real form of Eq.\ (\ref{eq:5.44}), this
expression can be simplified to
\begin{eqnarray}
A_i &=& \partial_i\phi[(\tilde{\chi}_1^{(1)})^2 +
(\tilde{\chi}_1^{(2)})^2]
\nonumber\\
&=& \frac{G_1^2}{2}\left[
\frac{1}{(\omega+\Omega^{(1)})(\omega+\Omega^{(1)}-2G_0Q)+
(G_0^2-G_1^2)Q^2}
\right. \nonumber \\ && \left.
+\frac{1}{(\omega+\Omega^{(2)})(\omega+\Omega^{(2)}+2G_0Q)+
(G_0^2-G_1^2)Q^2}
\right](-Q_2,Q_1)_i.
\label{eq:5.493}
\end{eqnarray}
In contrast to the previous model the vector potential is not
singular at the origin, but take the finite value
\begin{equation}
{\bf A}(0) = \frac{G_1^2}{4 \omega(0)^2}(-Q_2,Q_1).
\end{equation}

Turning to the off-diagonal matrix elements, we find by similar
calculations
\begin{eqnarray}
(A_i)_{11,0}&=&0,   \label{eq:5.494}  \\
(A_i)_{20,0}&=& \sqrt{2}\exp(-i\phi)[(\partial_i\phi)\psi_1^{(1)}
\tilde{\chi}_1^{(1)} +i(\psi_1^{(1)}\partial_i\tilde{\chi}_1^{(1)}
- \tilde{\chi}_1^{(1)}\partial_i\psi_1^{(1)}]
\nonumber\\ & = &
-\sqrt{2}\exp(-i\phi)\left(\psi_1^{(1)}\right)^2 f^{(1)}
\left[\partial_i\phi+i(\partial_iQ) \partial_Q \ln f^{(1)} \right]
\nonumber\\ & = &
\sqrt{2}\exp(-i\phi)
\frac{G_1[(\Omega^{(1)}+\omega)-G_0Q]}
{2[(\omega+\Omega^{(1)})(\omega+\Omega^{(1)}-2G_0Q)+(G_0^2-G_1^2)Q^2]}
\nonumber\\ && \times
\left[\frac{1}{Q}(-Q_2,Q_1)+i \frac{1}{Q}(Q_1,Q_2) Q\partial_Q
\ln f^{(1)} \right]
, \label{eq:5.495}\\
(A_i)_{02,0}&=& -\sqrt{2}\exp(-i\phi)
\frac{G_1[G_0Q+(\Omega^{(2)}+\omega)]}
{2[(\omega+\Omega^{(2)})(\omega+\Omega^{(2)}-2G_0Q)+(G_0^2-G_1^2)Q^2]}
\nonumber\\ && \times
\left[\frac{1}{Q}(-Q_2,Q_1)+i \frac{1}{Q}(Q_1,Q_2) Q\partial_Q
\ln f^{(2)} \right].
 \label{eq:5.496}
\end{eqnarray}
These off-diagonal potentials are also regular as $Q$ goes to zero,
for reasonably well behaved $\omega(Q)$.

\begin{figure}
%\epsf
\epsfysize=12cm
\centerline{\epsffile{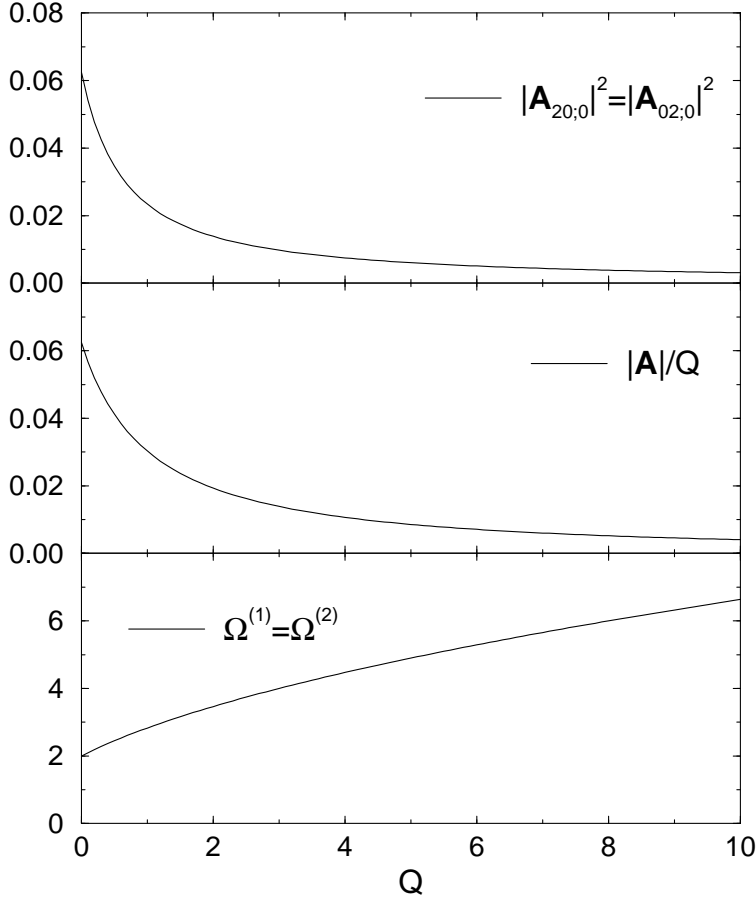}}
\caption{Parameters in the collective Hamiltonian for $G=1$,
$\omega_0=2$.
The lower panel shows the frequencies $\Omega^{(j)}$, the middle panel
shows the size of the diagonal Berry potential, and the
upper panel shows the square of the off-diagonal
Berry-potentials.\label{fig:G1om2}}
\end{figure}

We now study the numerical solution of the adiabatic Hamiltonian,
\begin{equation}
H_{C} = \frac{1}{2}[({\bf P}-{\bf A})^2 + |{\bf A}_{20,0}|^2 +
|{\bf A}_{02,0}|^2]
+\frac{1}{2}Q^2 + \frac{1}{2}[\Omega^{(1)}(Q)+\Omega^{(2)}(Q)],
\end{equation}
where we have once again made the choice $M=1,V(Q)=\frac{1}{2}Q^2$.
For simplicity we take $G_0=0$, and write $G_1=G$.
For this special choice $\Omega^{(1)}=\Omega^{(2)}=\Omega$ and
$(A_i)_{02,0}=-(A_i)_{20,0}$.
A simple form for $\omega(Q)$,
chosen such that  $\Omega$ is positive definite,
is
\begin{equation}
\omega(Q) = \omega_0+GQ.
\end{equation}
In Fig.~\ref{fig:G1om2} we show some of the relevant quantities
in the collective Hamiltonian for the choice $G=1$, $\omega_0=2$.
We  diagonalize the collective Hamiltonian by first going to
spherical coordinates, and use the fact that $m$ is conserved to
write down a radial equation. We then map the $Q$ values
from  the interval $[0,\infty)$ to the interval $[0,1]$. Finally
we make a finite difference approximation to the radial equation, and
solve
the approximate equation  by matrix diagonalization.

Solving the complete problem, without making the adiabatic
approximation
is somewhat involved. Since the Hamiltonian is invariant under the
interchange $a^\dagger_1 \leftrightarrow a^\dagger_2$, we introduce
new operators that are invariant under this parity transformation,
\begin{equation}
a^\dagger_\pm = \frac{1}{\sqrt 2} (a^\dagger_1 \pm a^\dagger_2).
\end{equation}
The interaction Hamiltonian takes the simple form
\begin{equation}
H_{int} = -G_0 (a^\dagger_+a_+ -a^\dagger_-a_-)
-\frac{G_1}{2} (Q_+[a_+^2-a_-^2] + Q_-[a_+^{\dagger2}-a_-^{\dagger2}]).
\label{eq:Hintnew}
\end{equation}
We now make the expansion
\begin{equation}
\Psi = \sum_{K,\kappa} \psi_{K,\kappa}(Q)
\frac{\left(a^\dagger_+\right)^{2K-2\kappa}
\left(a^\dagger_-\right)^{2\kappa}}
{\sqrt{(2K-2\kappa)!(2\kappa)!}}|0]
\end{equation}
(The even powers in this equation constitute the only combination
that includes the vacuum for the fast degrees of freedom.)

An approximate solution is obtained by limiting the sum over $K$,
while
summing over all allowed $\kappa$. At the same time we expand
$\psi_{K,\kappa}(Q)$ in  a finite number of
spherical harmonic oscillator eigenfunctions
$\langle Q| nm \rangle$.
We denote the $m$ value used in the expansion of
$\psi_{K,\kappa}(Q)$ by $m_{K,\kappa}$.
The interaction does not couple states of different $\kappa$ for
fixed $K$.
We thus find that
$m_{K,\kappa}=m_K=m_{K-1}-1$. Thus the value $m_0$ is
a constant of the motion, and can be used to specify different
solutions. This quantity  corresponds directly to the value of $m$
in the
collective Hamiltonian.
We have performed matrix diagonalizations for $\omega_0=2,10$ and
$G=1$.
We have used harmonic oscillator states up to principal
quantum number 70, and $K$ values up to 20.
(This corresponds to a 61,401 by 61,401 matrix.)
The resulting matrix,
which is very sparse, was diagonalized using a Lanczos algorithm.
The eigenvalues were  checked for convergence by comparing
a calculation with smaller cut-offs on $n$ and $K$ against one
with larger cutoffs. In tables \ref{tab:om2} and \ref{tab:om10}
we compare a few selected ground state energies of the complete
collective Hamiltonian with the exact solution. The splitting
between states with opposite values of $m$ is completely due
to the vector potential. For the case $\omega_0=2$, where the
adiabatic approximation we see that the size of the splitting
is very close to the exact value.  The difference is probably due
to non-adiabatic effects that can not be completely neglected
for  $\omega_0=2$. For the case $\omega_0=10$ the correspondence is
much closer, but the size of the
Berry vector potential is much smaller as well.

\begin{table}[htb]
\caption{A comparison between the exact numerical ground state
energies and those for the collective Hamiltonian. $\omega_0=2,G=1$,
and the values of $m$ are listed in the table.
\label{tab:om2}}
\begin{center}
\begin{tabular}{l|llll}
           &  $m=-5$ & $m =5$ & $m=-1$ & $m=1$ \\
\hline
exact      &  9.54715 & 9.73101 & 4.97309 & 5.03405 \\
collective &  9.56722 & 9.74656 & 4.98669 & 5.04375
\end{tabular}
\end{center}
\end{table}
\begin{table}[htb]
\caption{A comparison between the exact numerical ground state
energies and those for the collective Hamiltonian. $\omega_0=10,G=1$,
and the values of $m$ are listed in the table.\label{tab:om10}}
\begin{center}
\begin{tabular}{l|ll}
           &  $m=-5$ & $m =5$ \\
\hline
exact      &  18.0745 & 18.0918 \\
collective &  18.0750 & 18.0922
\end{tabular}
\end{center}
\end{table}

\section*{Acknowledgement}
This work was supported in part by the U.\ S.\ Department of Energy
under grant number 40264-5-25351.

\renewcommand{\theequation}
{\Alph{section}.\arabic{equation}}

\appendix
\setcounter{equation}{0}
\section{Scalar product and transformation of momentum operators}
In this appendix we provide the proof of Eq.\ (\ref{eq:2.17}), which
specifies how the momentum operator transforms under a general point
transformation.  It will be essential to recognize that this result
is tied to a choice of scalar product.  We suppose that the Hamiltonian
(\ref{eq:2.1}) is to be used in conjunction with the metric
\begin{equation}
\langle\Psi_a|\Psi_b\rangle\equiv\int d\xi^1\cdots d\xi^N
\Psi^{\ast}_a(\xi)
\Psi_b(\xi).                \label{eq:A.1}   \end{equation}
Thus                         \begin{equation}
\pi_\alpha\rightarrow -i(\partial/\partial\xi^\alpha).
\label{eq:A.2}            \end{equation}
Now carry out the point transformation (\ref{eq:2.2}) with
Jacobian $J$,
\begin{equation}    J=|\partial\xi^\alpha/\partial x^\beta|.
\label{eq:A.3}             \end{equation}
If we introduce a new wave function     \begin{equation}
\psi_a = J^{\half}\Psi_a ,   \label{eq:A.4}    \end {equation}
the metric is preserved in the sense        \begin{equation}
(\Psi_a,\Psi_b)=(\psi_a,\psi_b)=\int dx^1\cdots dx^N\psi^{\ast}_a(x)
\psi_b(x).       \label{eq:A.5}          \end{equation}
This is the metric that is associated with Eq.\ (\ref{eq:2.17}),
as we proceed to show.

With $d\xi^1\cdots d\xi^N \equiv [d\xi]$, we study     \begin{equation}
\int [d\xi]\Psi^{\ast}_a\pi_\alpha\Psi_b = \int [dx]\psi^{\ast}_a
J^{\half}\pi_\alpha J^{-\half}\psi_b.  \label{eq:A.6} \end{equation}
Thus we must show that            \begin{equation}
J^{\half}\pi_\alpha J^{-\half}=\half\{f^i_\alpha,(-i\partial/
\partial x^i)\}.     \label{eq:A.7}      \end{equation}
We have first                     \begin{equation}
J^{\half}\pi_\alpha J^{-\half}=J^{-\half}J\frac{\partial x^i}
{\partial\xi^\alpha}\frac{1}{i}\frac{\partial}{\partial x^i}J^{-\half}.
\label{eq:A.8}                    \end{equation}
For ease of notation, consider below a three-dimensional space, since
this will exhibit the general features.  By utilizing the equation,
equivalent to the definition of $J$ as a determinant, \begin{equation}
J\frac{\partial x^i}{\partial\xi^\alpha}=\half\epsilon_{ijk}
\epsilon_{\alpha\beta\gamma}\frac{\partial\xi^\beta}{\partial x^j}
\frac{\partial\xi^\gamma}{\partial x^k},
\label{eq:A.9}              \end{equation}
with $\epsilon_{ijk}$ the alternating symbol, we display the algebraic
manipulation leading to the desired result,      \begin{eqnarray}
J^{\half}\pi_\alpha J^{-\half}&=& \frac{1}{i}J^{-\half}\half
\epsilon_{ijk}\epsilon_{\alpha\beta\gamma}\frac{\partial\xi^\beta}
{\partial x^j}\frac{\partial\xi^\gamma}{\partial x^k}\frac{\partial}
{\partial x^i}J^{-\half}   \nonumber \\
&=& \frac{1}{2i}J^{-\half}\{\frac{\partial}{\partial x^i},\half
\epsilon_{ijk}\epsilon_{\alpha\beta\gamma}\frac{\partial\xi^\beta}
{\partial x^j}\frac{\partial\xi^\gamma}{\partial x^k}\}
J^{-\half}   \nonumber \\
&=& \frac{1}{2i}J^{-\half}\{\frac{\partial}{\partial x^i}\half
\epsilon_{ijk}\epsilon_{\alpha\beta\gamma}\frac{\partial\xi^\beta}
{\partial x^j}\frac{\partial\xi^\gamma}{\partial x^k}
J^{-\half}\} \nonumber \\
&&+\frac{1}{2i}\{J^{-\half}\half
\epsilon_{ijk}\epsilon_{\alpha\beta\gamma}\frac{\partial\xi^\beta}
{\partial x^j}\frac{\partial\xi^\gamma}{\partial x^k}\frac{\partial}
{\partial x^i}\}J^{-\half}  \label{eq:A.10}     \end{eqnarray}
By an application of (\ref{eq:A.9}), this becomes  \begin{equation}
\frac{1}{2i}J^{-\half}\frac{\partial}{\partial x^i}(J^{\half}\frac
{\partial x^i}{\partial\xi^\alpha})
+\frac{1}{2i}J^{\half}\frac{\partial x^i}{\partial\xi^\alpha})
\frac{\partial}{\partial x^i}J^{-\half} = \frac{1}{2i}\{\frac{\partial
x^i}{\partial\xi^\alpha},\frac{\partial}{\partial x^i}\},
\label{eq:A.11}            \end{equation}
the result sought.

\setcounter{equation}{0}
\section{Potential energy for the fast variables}

In terms of the original coordinates, let us consider the change in the
potential energy between two neighboring points, $\xi$ and $\xi +\delta
\xi$.  To second order in $\delta\xi$ we have the usual terms of a
Taylor expansion,               \begin{eqnarray}
\Delta V &=& V(\xi +\delta\xi) - V(\xi)   \nonumber \\
&=&  V_{,\alpha}\delta\xi^\alpha +\half V_{\alpha\beta}\delta\xi^\alpha
\delta\xi^\beta,   \label{eq:B.1}   \end{eqnarray}
that now appears most unsatisfactory, since $\Delta V$ is a scalar,
but the second term of (\ref{eq:B.1}) contains the ordinary rather than
the covariant second derivative.  This defect is removed by replacing
the quantity $\delta\xi^\alpha$ by substitution from the relation
                                   \begin{equation}
d\xi^\alpha = \delta\xi^\alpha +\half\Gamma^\alpha_{\beta\gamma}
\delta\xi^\beta\delta\xi^\gamma,   \label{eq:B.2}  \end{equation}
that contains the Christoffel symbol,    \begin{equation}
\Gamma^\alpha_{\beta\gamma} =\half \tilde{B}^{\alpha\delta}
(\tilde{B}_{\delta\beta,\gamma} + \tilde{B}_{\delta\gamma,\beta}
-\tilde{B}_{\beta\gamma,\delta}).   \label{eq:B.3}  \end{equation}
We thus obtain the form                  \begin{equation}
\Delta V  =  V_{,\alpha}d\xi^\alpha +\half V_{;\alpha\beta}
d\xi^\alpha d\xi^\beta,    \label{eq:B.4}  \end{equation}
where the second term contains the covariant derivative
\begin{equation}
V_{;\alpha\beta} = V_{,\alpha\beta} -\Gamma^\delta_{\alpha\beta}
V_\delta.
                             \label{eq:B.5}
\end{equation}
It is now apparent from (\ref{eq:B.4}) that $d\xi^\alpha$ are the
components of a vector, and therefore transformation to any alternative
set of coordinates such as the $q^\mu$ is standard.

This allows us to
calculate the quantity $V^{(2)}$ of Eq.\ (2.10) from given dynamical
quantities, provided we can define a complete set of coordinate axes
at each point of the decoupled manifold.  This is done in two stages.
In the first stage, the basis vectors at each point of the tangent
space
to the collective submanifold, for example
the set $f^i_\alpha$, are determined by the algorithm that discovers
the collective submanifold. A set of basis vectors $f^a_\alpha$,
orthogonal to the tangent space is then determined (non-uniquely)
by the requirement that these be orthogonal to the $f^i_\alpha$ and to
each other with respect to the metric $\tilde{B}^{\alpha\beta}$.
The basis vectors orthogonal to the tangent space are precisely the
elements needed to compute the Hamiltonian of the fast variables.

\end{document}